\documentstyle[aps,manuscript]{revtex}
\newcommand{\be}{\begin{equation}}
\newcommand{\ee}{\end{equation}}

\def\simlt{\lower.5ex\hbox{\ltsima}}
\def\gtsima{$\; \buildrel > \over \sim \;$}
\def\simgt{\lower.5ex\hbox{\gtsima}}

\def\simlt{\lower.5ex\hbox{\ltsima}}
\def\gtsima{$\; \buildrel > \over \sim \;$}
\def\simgt{\lower.5ex\hbox{\gtsima}}

\def\cm{{\rm\,cm}}

\def\ergcm2{\ {\rm erg~cm^{-2} }}
\def\ergscm2{\ {\rm erg~s^{-1}~cm^{-2} }}

\def\cm2s{\ cm^2 ~s^{-1} }

\def\s{\ifmmode \widetilde \else \~\fi}
\def\={\overline}

\def\spose#1{\hbox to 0pt{#1\hss}}

\def\lta{\mathrel{\spose{\lower 3pt\hbox{$\mathchar"218$}}
     \raise 2.0pt\hbox{$\mathchar"13C$}}}
\def\gta{\mathrel{\spose{\lower 3pt\hbox{$\mathchar"218$}}
     \raise 2.0pt\hbox{$\mathchar"13E$}}}
\def\mincir{\ \raise -2.truept\hbox{\rlap{\hbox{$\sim$}}\raise5.truept  
\hbox{$<$}\ }}                                                          %
\def\magcir{\ \raise -2.truept\hbox{\rlap{\hbox{$\sim$}}\raise5.truept  %
\hbox{$>$}\ }}                                                          %
\def\simlt{\ \raise -2.truept\hbox{\rlap{\hbox{$\sim$}}\raise5.truept   
\hbox{$<$}\ }}                                                          %
\def\simgt{\ \raise -2.truept\hbox{\rlap{\hbox{$\sim$}}\raise5.truept   %
\hbox{$>$}\ }}                                                          %
\def\newline{\par\noindent}

\def\s-z{S-Z}

\begin{document}

\title{Probing The Structure of Space-Time with Cosmic Rays}

\author{Roberto Aloisio$^{1,4}$, Pasquale Blasi$^2$, Piera L. Ghia$^{3,5}$,
Aurelio F. Grillo$^4$}
\address{$^1$ Dipartimento di Fisica, Universit\`a di L'Aquila,
Via Vetoio, 67100 Coppito (L'Aquila) - Italy} 
\address{$^2$ NASA/Fermilab Astrophysics Group, Fermi National Accelerator
Laboratory, Box 500, Batavia, IL 60510-0500}
\address{$^3$ Istituto di Cosmo-Geofisica del CNR, Corso Fiume 4, 
10133, Torino, Italy}
\address{$^4$ Laboratori Nazionali del Gran Sasso, INFN, SS. 17bis, Assergi
(L'Aquila) - Italy}
\address{$^5$ Istituto Nazionale di Fisica Nucleare, Via P. Giuria 1, 10125,
Torino, Italy}

\maketitle

\begin{abstract}
The study of the interactions of Cosmic Rays (CR's) with 
universal diffuse background radiation can provide very 
stringent tests of the validity of Special Relativity.
The interactions we consider are the ones characterized by 
well defined energy thresholds whose energy position can be predicted 
on the basis of special relativity. We argue that the experimental 
confirmation of the existence of these thresholds can in principle 
put very stringent limits on the scale where special 
relativity and/or continuity of space-time may possibly break down. 
\end{abstract}

\section{Introduction}
\label{intro.sec}



The study of interactions of
high energy Cosmic Rays (nucleons and gamma rays) can provide severe 
tests of the validity of Special Relativity (SR).
Indeed, SR is at the very base of our theories for the 
description of the Universe, and its validity is generally not 
questioned. However, several attempts have been made to put under scrutiny 
the postulates that SR relies upon, like the constancy of the velocity
of light, and these studies have contributed accurate limits on 
possible violations (see e.g. \cite{colgla} and references therein).

Although it is very well proven that SR and the underlying Lorentz
invariance (LI) provide a suitable framework for our low energy effective 
theories, it is not clear whether more ambitious theories, aiming to a 
global description of our World, including gravity, are or need to be
Lorentz invariant. 
Theoretically, the need for a unified theory of gravity 
has led to several models, some of which automatically imply the 
breaking of LI. This is qualitatively understandable since the very concept 
of continuous space-time is likely to be profoundly changed by quantum 
gravitational effects \cite{wheeler57}. 
In this prospective we think it is worth keeping 
an open mind with respect to the validity of what we define as 
fundamental theories, and always put them under experimental
scrutiny.

This attitude appears even more justified at the present time: 
observationally, high 
energy astrophysics is providing a range of opportunities to probe
energies ({\it i.e.} Lorentz factors, or speeds) 
much larger than the ones ever obtained, 
or obtainable  in the future, 
in accelerator experiments.
In particular we concentrate our attention on the interactions of 
cosmic gamma rays and nucleons  with some type of universal photon 
background, {\it i.e.} the cosmic microwave  (CMB), the 
far infrared (FIRB), and the radio background. 
These are the interactions responsible for gamma ray
absorption from distant sources 
(through production of $e^+ e^-$) \cite{niki,gold,gould}, 
that should appear as a cutoff in the
gamma ray spectrum, and of the well known Greisen-Zatsepin-Kuzmin (GZK)
cutoff \cite{greis,kz} due to the photopion production in collisions
of ultra high energy cosmic rays (UHECRs) off the CMB photons.
All these processes have the same general structure: in a Lorentz 
invariant picture, these reactions would be examples of very low energy 
processes in the center of momentum, that appear boosted to very large
Lorentz factors in the laboratory frame. Testing the presence of the
thresholds for these processes is therefore a test of SR up to the
large Lorentz factors in the boost.

Our approach is entirely phenomenological, and reasonably model
independent. We do not propose, if not as examples, any specific 
model. We show that when a very general form of LI 
violation
\footnote{Clearly, breaking LI can have several implications, 
as for instance the existence of a 
preferred reference frame, which is the one in which all the
calculations need to be carried out, since breaking LI also 
invalidates the transformations that allow us to change reference
frame. 
This reference frame could be identified with the one
comoving with the expansion of the universe (no peculiar motion), 
in which the microwave background is 
completely isotropic. In fact, there is only one frame with this 
property, being all other frames experiencing the dipole anisotropy,
and therefore distinguishable.} 
is explicitly allowed 
in the dispersion 
relation between energy and momentum of particles, 
the threshold momenta 
for some of these reactions may drastically change or even 
become unphysical ({\it i.e.} the process does not occur),
unless the LI violation
is introduced at a length scale much smaller 
than the Planck scale. 
A similar approach, although more specific, was proposed in \cite{kir}
(see Sect. 3).

The aim of the present paper is to discuss the possibility that future 
CR experiments could test Special Relativity and/or the structure of 
space-time \cite{mav1}. In our opinion, the (lack of) 
knowledge  of the sources of
the CR's under consideration does not allow us to support the idea
\cite{colgla,lgm,klu,kifu} 
that the present experimental situation gives evidence for 
violation of LI or of continuity of space-time. Rather we stress that 
a verification of the existence of the quoted thresholds would 
entail a lower limit on the mass scale of effective
LI violations. 

The paper is planned as follows: in section 2 we discuss our parametrization
of the LI violations, and discuss briefly some models in which these 
parametrizations hold; in section 3 we apply our calculations to the case of 
pair production and photopion production in high energy cosmic ray
interactions. We conclude in section 4.

\section{Breaking of Lorentz Invariance}
\label{format.sec}

Lorentz invariance implies that the modulus of any four-vector is unchanged
when changing reference frame; for instance for the
four-momentum of a particle we have (we always put $c=\hbar=1$) 
$$P_{\mu}P^{\mu}=
E^2-p^2=const=m^2$$ 
$m$ being the invariant particle mass and $p=|\vec p|$. 

Violations of Lorentz invariance will therefore affect in general the 
{\it dispersion relation} above. 
Without referring to any specific model (later we will
discuss explicit implementations), we write a  modified 
dispersion relation obeying the following postulates:
\begin{enumerate}
\item Violations are a {\it high energy} effect, {\it i.e.} they vanish
at small momenta.
\item Violations are universal, {\it i.e.} do not depend on the particle
type, if not (possibly) through the particle mass.
\item Rotation invariance remains exact.
\end{enumerate}

Clearly requirements 2 and 3 may be relaxed and in fact there are 
examples in this sense \cite{colgla,bog}.
We write then the modified dispersion relation as follows:

\begin{equation}
E^2-p^2-m^2= p^2 f({p\over M})
+ m^2 g({p\over M})
\label{eq:disp}
\end{equation}

where the mass $M$ parametrizes the violation of Lorentz invariance (or an 
essential discreteness of space-time, as for instance suggested by 
some models of quantum gravity 
(e.g. \cite{wheeler57} and \cite{ell1,mav2,loop})).
Even in the framework defined above this is not the most general violation
term. However, in the regime we are interested in, 
$m\ll p\ll M$, the left hand side of eq. (\ref{eq:disp}) is small
compared to $p^2$ and $E^2$ so that the other 
possible terms we can write 
(containing for instance $E$) differ from those in eq. (\ref{eq:disp})
by higher order corrections
\footnote{
Also, terms proportional to $mp$ can be added. They however do not 
modify the general framework so we will not include them for sake of
clarity.}.

Since $p/M\ll 1$ the functions $f$ and $g$ can be Taylor-expanded to give:
$$
E^2-p^2-m^2=
p^2(f(0)+f'(0){p\over M}+f''(0){p^2\over M^2}+...)+$$
\begin{equation}
m^2(g(0)+g'(0){p\over M}+g''(0){p^2\over M^2}+...)
\end{equation}
In the limit $M \to \infty$ one must recover the Lorentz invariant
dispersion relation, so $f(0)=g(0)=0$; moreover the linear term 
might be absent, as we will see later; if it is present, the quadratic 
term is negligible at the momenta we consider. 
The coefficients in front of the first and second derivatives are 
left unconstrained.
However, if $|f'(0)|,|g'(0)|\gg 1$ or $|f''(0)|,|g''(0)|\gg 1$, 
then the functions 
$f$ and $g$ would 
be strongly varying and possibly oscillating, which would be 
unphysical. Moreover, in cases where the calculations of the
LI breaking can be carried out explicitly (see section 2A), these
coefficients turn out to be of order unity (the same motivation justifies
the assumption that these coefficients are not much smaller than unity). 
Therefore we decided to adopt
here a phenomenological approach and assume that possible small deviations
from unity are embedded in the mass scale $M$. The validity of this 
approach will be checked {\it a posteriori}.

We are thus led to the following classification of Lorentz non-invariant 
dispersion relations: 

\begin{equation}
I_{\pm}:\quad \quad E^2-p^2 \approx m^2
\pm {p^3\over M} \quad (\pm {m^2 p\over M})
\label{eq:Ipm}
\end{equation}

\begin{equation}
II_{\pm}:\quad \quad E^2-p^2 \approx m^2
\pm {p^4\over M^2} \quad (\pm {m^2 p^2 \over M^2})
\label{eq:IIpm}
\end{equation}

where $I$ ($II$) stands for first (second) order modification and the 
terms in parenthesis come from the expansion of $g$.

We can define a critical momentum $p_c$ where the correction,
for massive particles, 
equals $m^2$, which is the momentum for which we expect that
deviations from normal relativistic kinematics become relevant. 
We have $p_c \approx 2 \times 10^{15}$ eV ($\approx 10^{13}$ eV)
for protons (electrons) in the case $I_{\pm}$ and
$p_c \approx 3 \times 10^{18}$ eV ($\approx 10^{17}$ eV) for
protons (electrons) in the case $II_{\pm}$.
In the cases in parenthesis in
eqs. (\ref{eq:Ipm}) and (\ref{eq:IIpm}) we have always 
$p_c \approx M$, so these modifications do
not lead to observable consequences at the energies we are interested in;
we will therefore put
$g({p \over M})=0$ in the following. We will come back to this point 
when we will describe specific examples of Lorentz violating theories
\cite{kir}.
It is clear that the values of $p_c$ given above are calculated in
a specific frame; 
in fact, generally, violations of LI imply the existence of a 
privileged frame.
We choose this frame to be the one comoving 
with the expansion of the universe (we name it ``universal
frame''), and we argue that this choice 
is in fact not arbitrary: this is the only possible frame where the
microwave background is isotropic (the same holds for the other
backgrounds, provided the sources are homogeneously and 
isotropically distributed). Moreover, neglecting the proper motion
of the Earth, this is the reference frame in which we live
and measure the thresholds for physical processes. On a more 
practical ground, it is worth noticing that giving up LI, 
the Lorentz transformations do not correctly give the transformations
laws of energy and momentum between different frames
although in principle it is possible to write
modified transformations 
of energy and momentum (see for instance \cite{russo})
which reproduce, at least at the 
perturbative level chosen,  the dispersion 
relations above.
In this case, however, there is much more arbitrariness than in modifying
the dispersion relation and we do not pursue 
this approach in this paper. It 
is worth noting however that the Lorentz Transformations 
are derived in a LI theory
by the requirement of the invariance of a fundamental interval, so
in a sense a modification of the dispersion relation is more
fundamental.

There is a further, important point to discuss before computing
particle production thresholds. In fact, when one gives up 
relativistic invariance, energy-momentum conservation is
not anymore guaranteed, and relativistic quantum field theory
may fail, so there is no guidance in deriving cross sections.
However, our point of view here is to derive the consequences
of experimental {\it verification} of the existence of the 
particle thresholds, so we will {\it assume} in the following
exact energy-momentum conservation {\it and} relativistic
dynamics in the preferential frame. 

\subsection{Models}

In the following we will briefly discuss some models in which 
modified dispersion relations are actually obtained.
It has been argued on very general grounds that 
quantum gravity effects do modify particle propagation at scales close to the
Planck scale. In particular, modified uncertainty
relations and existence of a $minimum$ proper 
length (see e.g. \cite{garay} and references therein) 
(implying modifications of Lorentz transformations),
and light-cone fluctuations (see e.g. \cite{ford} and references therein)
are fairly general implications of quantum gravity and modify the 
dispersion relations.

An approach where a fundamental mass/momentum scale $M$ is introduced to 
parametrize the deviations from LI can be pursued by writing  
the commutators between (space) boosts and translation generators of the
Poincare' algebra in a modified form \cite{luk1}, leading to

\begin{itemize}
\item{$II_+ \qquad m^2=M^2\sin^2(\frac{E}{M})-p^2 \sim
E^2-p^2-\frac{E^4}{12M^2}$}
\item{$II_- \qquad m^2=M^2\sinh^2(\frac{E}{M})-p^2 \sim
E^2-p^2+\frac{E^4}{3M^2}$}
\end{itemize}

A theory with both exact conservation of energy and momentum and LI 
violations can be obtained if it is possible to construct a local theory 
which is symmetric under the modified Poincar\'e group \cite{luk4}

Models giving the  $I_{\pm}$ dispersion relations can be derived from 
some quantum gravity approach. 
The basic idea is that quantum fluctuations of gravity
cause, at a scale of Plank mass, the vacuum to behave like a stochastic medium
and this introduces non-zero, energy dependent  non-diagonal terms in the
metric \cite{ell1}, 

\begin{itemize}
\item{$I_+$}: ~~~$g_{\mu\nu}\to g_{\mu\nu}+h_{0i} \qquad h_{0i}=U_i 
\quad \vec{U}=-\frac{E}{M}\hat{p}$ 
\item{$I_-$}: ~~~$g_{\mu\nu}\to g_{\mu\nu}+h_{0i} \qquad h_{0i}=U_i 
\quad \vec{U}=\frac{E}{M}\hat{p}$ ,
\end{itemize}
with consequent modifications of the dispersion relations.
Similar modifications may hold in brane models with large (or even infinite)
extra dimensions and $TeV$ scale quantum gravity \cite{elltev},
where in some cases Poincare' invariance is broken explicitely 
\cite{tilting}.

In the models presented above, quantum gravity effects are described by 
fluctuations around a flat background metric. An entirely different
approach is followed in the loop approach to quantum gravity 
(see e.g \cite{loop} and references therein)
in which the geometry itself emerges non perturbatively. It has been
shown that this leads to an essential discretization of space, and
to modifications of the dispersion relations which fall in the class
$I_{\pm}$ if parity is broken; the violation might be milder (or absent)
in parity conserving models.

We finish  quoting a pioneering approach by Kirshnitz and Chechin \cite{kir},
stimulated by the appearance of the papers by Kuzmin and Zatsepin \cite{kz}, 
and Greisen \cite{greis}. It is a classical approach, in which the free
lagrangian of a (classical) point particle is modified.
The theory is defined replacing the  pseudo-Euclidean  
space-time of SR with a Finslerian space \cite{run}.
In this approach  the dispersion relation becomes

\begin{equation}
m^2=[1+\bar g(\alpha\frac{p^2}{E^2})](E^2-p^2),
\end{equation}

where $m$ is the particle mass,  and $\bar g$ is a homogeneous 
function of the dimensionless parameter $\alpha$ that parametrizes the 
LI violations, $\alpha=\frac{m^2}{M^2}$ in terms of the scale M.
This gives rise to a mild  violation, disappearing for massless particles.
With an appropriate choice of the function $\bar g$, this gives rise 
to the terms in parenthesis in eqs. (\ref{eq:Ipm}) and (\ref{eq:IIpm}).

\section{Threshold calculations with modified dispersion relations}

In this section we describe the calculation of the kinematic thresholds for
some processes in the framework
of the modified dispersion relations between energy and momentum introduced 
in the previous section.
We choose two processes which are of astrophysical relevance and that will be
accessible to next generation cosmic ray experiments: pair production in
photon-photon scattering and photopion production in nucleon-gamma scattering.
The first process is responsible for the absorption of high energy gamma rays 
from distant sources, while the second is responsible for the well known 
(and currently unobserved) GZK cutoff.

\subsection{ $e^+ e^-$ production in photon-photon interactions}

The process under investigation is $ \gamma \gamma \to e^+ e^-$.
The energy of the background photons is taken as $\omega=|k|$ 
(where $k$ is the photon momentum) since 
at the typical background momenta 
(FIRB: $\omega \approx 0.01$ eV,
CMB: $\approx 6 \times 10^{-4}$ eV, and radio: $\approx 4\times 10^{-9}$
eV for the peak of the corresponding radiance distributions)  
the corrections are entirely negligible. 
We compute the threshold, assuming that the CR particle and the
background photon collide head-on, and the final particles are collinear. 
This is indeed not an arbitrary configuration, but the one that 
provides the minimum energy for which the process can occur in
the universal frame.

Rotational invariance then implies that in this reaction the momenta of
final particles are equal, and that particles move in the direction of 
the primary, so that the problem is one-dimensional; the energy of the
background photon is assumed to be equal to the average for the photon 
background in consideration.

Let then ($E,p$) be the energy  and (modulus of) the momentum of the
incident photon. Writing the relations of conservation of energy and
momentum in the laboratory frame, and using the modified dispersion relation
between energy and momentum, after some trivial algebra, 
and neglecting sub-leading terms in the range of momenta 
we are considering,
we get  the following general 
equations for the threshold:

\begin{equation}
I_{\pm}:\quad  \pm\alpha_I x^3+x-1=0,
\label{eq:eqIpm}
\end{equation}
\begin{equation}
II_{\pm}:\quad  \pm\alpha_{II} x^4+x-1=0,
\label{eq:eqIIpm}
\end{equation}

where $x=p_{th}/p_0$, $p_0=m^2/\omega$ is the threshold for $M\to \infty$

({\it i.e.} the usual threshold, 
$\approx 3\times 10^{13}$ eV for the FIRB as background,
$\approx 5 \times 10^{14}$ eV for the CMB, and finally 
$\approx 6 \times 10^{19}$ for the radio background), 
$m$ the electron mass, and:

\begin{equation}
\alpha_I=\frac{p_0^3}{8m^2 M}; \quad \quad
\alpha_{II}=\frac{3p_0^4}{16m^2 M^2}.
\end{equation}

The modified thresholds are the positive real solutions of the 
equations above, and in particular, if more than one positive solution 
is present, the one which goes to $x=1$ as $M \to \infty$. 
In table \ref{Table1} we report the solutions (if any) of the eqs.
(\ref{eq:eqIpm}) and (\ref{eq:eqIIpm}) in the case $M = M_P$, 
reasonable in quantum gravity models, for
the infrared, 
microwave,
and radio background.

The general feature of these solutions is that a {\it positive}
modification of the dispersion relation tends to move the thresholds
towards lower momentum values; {\it negative} modifications tend to 
lead to {\it complex} solutions, meaning that the kinematics of the process
becomes forbidden, {\it i.e.} the threshold disappears.
Case II is less predictive, as it is obvious being the modification
of second order in the (small) quantity $p/M$. 
Clearly a more detailed calculation, accounting for the integration over
different scattering angles and over the whole spectrum of the photon
background, could give slightly different results, not changing however 
our basic findings.

To assess the role of CR
experiments to detect violations of LI we now {\it assume}
that in present or future experiments it will be possible to isolate
the effect of pair production on different photon backgrounds,
by measuring the respective thresholds, with
an experimental uncertainty on the measurement of the threshold
energy of $\approx 100 \%$. This is a conservative 
(and probably pessimistic) approach, unless 
possibly for the radio case. We therefore derive a lower limit 
on $M (\ne M_P)$, 
which we report in table \ref{Table2}.

Note that all these limits 
are more stringent than the few limits on $M$ already
obtained (see for instance \cite{biller}). Exceptions to this statement
come from the cases $II_{\pm}$ for the IR and 
microwave backgrounds, where
the limits on $M$ are appreciably smaller that $M_P$, leaving more 
room for non Lorentz invariant theories.

We now discuss 
the present situation of VHE/UHE $\gamma$-ray
astronomy in view of the possibility of testing LI. 
In doing this it is important to keep in mind
the quite different observational situation existing for the 
CMB versus the IR and Radio backgrounds: while the former is
determined with
very high accuracy, the IR and Radio backgrounds are very poorly 
known and difficult
to access through observational investigation, 
mainly due to the emission and/or absorption processes in our
Galaxy.

Observationally, the infrared gamma-ray cutoff seems
to be the one more easily approachable. Indeed signs of a cutoff in
the TeV spectrum of a few blazars seem to be already present 
(\cite{whipple,hegra,vvv,stecker,coppia}) 
and helped to impose some constraints on the extragalactic 
far-infrared background (FIRB). 
Unfortunately, as it is clear from table \ref{Table1}, the
case of interaction with the FIRB is not very predictive, in the 
sense that the threshold is not appreciably changed with respect to the 
case of a Lorentz-invariant theory, for most of the possible parametrizations
of LI breaking (using $M=M_P$). The case $I_-$ is an exception: 
in this case there
is no acceptable solution of the equations that define the threshold, 
which means that the process is not possible at all. In other words, in
this case no cutoff should be seen in astrophysical observations 
because photons can come unattenuated from all distances. Viceversa, if
a cutoff is observed, then a lower limit on $M$ can be imposed.

As far as the photon absorption
on the CMB radiation is concerned, the 
experimental situation is more uncertain, due to the higher 
energy of the $\gamma$-rays involved
(although there might be suggestion of attenuation due to 
CMB, see e.g. \cite{casa,eastop}).
This is theoretically more restrictive:  in the
scenario $I_+$ the threshold is moved to smaller values. For $M=M_P$,
photons with energy as small as $\sim 25$ TeV from a distant
source should be absorbed.
In order for the threshold to be unchanged, the scale $M$ must 
exceed $\sim 800 M_P$. 
In the scenario $I_-$ the process becomes not allowed, while in the
other cases the threshold remains unchanged.   

The case of scattering off the radio background is the most 
interesting. We predict, for $M=M_P$, 
either no threshold (cases $I_-$ and $II_-$)
or thresholds which are much smaller that the canonical ones. 
This could be of great importance for top-down models of UHECRs,
where gamma rays are supposed to have an important role in the
composition (e.g. see \cite{bbv}). 
In particular the proton/gamma ratio is determined 
by the interaction of the gamma rays with the radio background at
frequencies smaller than a few MHz.
Unfortunately, as mentioned above, 
the radio background at these frequencies is extremely
uncertain, and not accessible to any direct measurement, due to the
strong free-free absorption in the disc and halo of our own Galaxy.

\subsection{Pion photoproduction in UHECR interactions: the GZK cutoff} 

This case is slightly more involved since the masses of the final particles
are different, so that even for a rotation invariant modification, final
momenta are not in the same ratio as the masses. However we checked that
assuming the ratio of final momenta as in the Lorentz invariant 
theory, we introduce only higher 
order corrections. Using this prescription we obtain the following
two equations for the threshold, with the same symbols used in the 
previous section:

\begin{equation}
I_{\pm}:\quad  \pm\alpha_I x^3+x-1=0
\end{equation}
\begin{equation}
II_{\pm}:\quad  \pm\alpha_{II} x^4+x-1=0
\end{equation}

where 
$x=\frac{p_{th}}{p_0}$ and $p_0=\frac{m_{\pi}^2+2m_{\pi}m_p}{4\omega}$ 
is the conventional threshold (for $M\to \infty$), 
$m_p$ is the proton (neutron) mass, and 
$m_{\pi}$ is the pion mass. The coefficients $\alpha_{I,II}$ are defined as
follows:

\begin{equation}
\alpha_I=\frac{2p_0^3}{(m_{\pi}^2+2m_{\pi}m_p)M}
\frac{m_\pi m_p}{(m_\pi+m_p)^2}; 
\end{equation}
$$
\alpha_{II}=\frac{3p_0^4}{(m_{\pi}^2+2m_{\pi}m_p)M^2}
\frac{m_\pi m_p}{(m_\pi+m_p)^2}.
$$ \\

Fixing $M=M_P$ we have the solutions reported in table \ref{Table3}.

The general trend of the solutions is the same as
in the case of pair production. Here however we can easily see that 
the consequences are even more evident: in all cases either the threshold
disappears and the process becomes forbidden (cases $I_-$ and $II_-$), or 
the threshold is appreciably lowered (cases $I_+$ and $II_+$). 
For $I_-$ or $II_-$ parametrizations, the GZK cutoff is completely
washed out and particles (nucleons) should be able to reach us from
any distance. In the other cases the threshold falls in a region that 
is easy to discard even on the basis of the present data, unless the
scale of breaking of LI is $\sim 3\times 10^{13}$ ($\sim 500$) 
times larger than the Planck mass for the case $I_+$ ($II_+$). 

As in the previous section, if one assumes that the apparent absence of
the GZK cut-off is an incidental fact, possibly due to our ignorance of the
sources of CR's at these energies, and that CR experiments may find
some remnant of the cut-off, we can derive limits on the parameters of 
LI violations. These limits are
reported in table \ref{Table3}, again assuming a $100 \%$ error on the
location of the cut-off. From the values reported 
it is clear that
UHECR experiments soon to be available will provide a powerful tool to
explore the very-small-scale structure of space-time. 

\section{Conclusions}

Special Relativity is at the base of our understanding of the physical
world. However any physical concept should always be put under 
stringent experimental verification. This is 
particularly true in the case of SR, specially in connection with 
the quest for a fundamental theory of Nature, which includes gravity.
Any such theory will imply a full knowledge
of what the vacuum really is at the level where quantum fluctuations
build the space-time, possibly modelling it with a non-trivial 
geometry. Some first attempts to reach this goal seem to suggest
that Lorentz invariance, one of the building blocks
of our current low energy theories, could be broken at extremely
high energy. 

Lacking a true theory of quantum gravity, the most we can do at
present is to adopt a phenomenological approach and ask ourselves
whether there is any probe that can be used to check the validity 
of SR at very high energy. This is precisely the approach adopted
in this paper, where high energy cosmic rays (nucleons and gamma 
rays) have been used as probes. 
We found that the thresholds for pair production and photopion
production off some universal photon backgrounds,
which are experimentally accessible or will be in the
next generation of cosmic ray experiments,  are often
profoundly affected
by the possibility of breaking LI at supra-Planck scales. As a 
consequence, any quantum gravity theory in which LI is a casualty
must face the cosmic ray bound in order to be viable.  

Our calculations were based on a perturbative but quite general 
modification of the dispersion relation between energy and momentum
of a particle. This modification affects in a fundamental way 
the calculation of the thresholds for pair production and photopion
production, making these processes forbidden in some cases, or 
lowering the thresholds to questionable values in others. Building
on this approach we proposed a procedure to obtain 
very strong constraints on the
energy (or length) scale at which a possible LI violation could occur.

This procedure relies upon the possibility that
astrophysical observations will be able in the near
future to find evidence for these processes, through cutoffs in
the TeV spectra of distant blazars or/and through the discovery
of possible consequences
of the GZK cutoff in the spectrum of UHECRs or/and through 
studies of the composition of UHECRs. With a few exceptions, 
these limits are all higher than the Planck scale, which is
an indication that breaking
LI is not necessarily a safe ingredient of unified theories that 
require it.

We stress that our approach is purely kinematical and no dynamical
effect is considered. Moreover, we assume
perfect energy-momentum conservation in order to compute the thresholds.
While these are clearly important issues for any calculation aiming
to {\it predict} values for specific particle production thresholds,
and consequently absorption cutoffs in CR spectra, our approach is 
rather the opposite: we want to discuss the consequences of a
possible experimental verification of the presence of
particle production thresholds. If a given absorption threshold 
is experimentally detected, discarding the possibility of miraculous 
compensations between relativity violations, non conservation of 
energy-momentum 
and non-relativistically invariant dynamics, we are forced to
conclude that possible violations of LI are smaller than
the sensitivity of the experiment, and, to make this statement 
quantitative, we {\it choose} to use a purely kinematical, 
energy-momentum conserving parametrization. 

Although a very tempting possibility, we do not believe that the current 
experimental situation allows us to draw 
definitive statements : recent observations of
Markarian 501 might suggest the presence of a cutoff in the TeV region,
but the unaffected spectrum is not known well enough to exclude that
the observed effect is the artifact of a cutoff in the production 
spectrum. Hopefully the situation will improve with the next generation
gamma ray detectors, complemented by neutrino and X-ray detectors that could
clarify the origin of the gamma ray emission. 
Concerning $\gamma-$rays absorption on CMB photons, experimental data are
more scarce, and the situation is not likely to improve in the 
near future.

Pair production on the universal radio background becomes relevant for
ultra-high-energy gamma rays, usually produced in top-down models of
UHECRs \cite{bbv}. The presence or absence of a threshold for this 
process strongly
affects our predictions of the fluxes at the Earth: although very uncertain,
the radio background should allow typical gamma ray pathlengths (in a 
Lorentz invariant world) of the order of $2-20$ Mpc \cite{pb}.  
If a violation of LI made the pair production kinematically 
forbidden, gamma rays could reach us from any distance and 
contribute an enormous flux of particles above the GZK cutoff.

As for UHECRs, although the (small) number of events \cite{tak,hires}
above 
$\sim 5\times 10^{19}$ eV seems already incompatible with the
presence of the GZK cutoff, the lack of knowledge of the sources
does not allow any firm statement on the propagation of primaries 
in the Universe.  The situation will 
improve dramatically in the near future, as new experiments like
HiRes \cite{hiresidea} and Auger \cite{auger} 
will collect data and reliable measurements on
observables different from the energy spectrum (anisotropy, clustering)
will be available with reasonable statistics.  

Our conclusions can be summarized in the following points: 1) it is
a 40-years old idea that the incompleteness of the present theories 
could rely upon our ignorance of the vacuum at very high energies.
When we will have that knowledge we will probably have a quantum theory 
of gravity; 2) several attempts to quantize gravity have naturally led
to the requirement of violations of the LI; 3) experiments on cosmic 
rays can represent the most easily approachable tool to probe the structure 
of space-time on the very small scales; 4) LI violations affect the
thresholds for elementary processes relevant for cosmic ray
astrophysics and can be observationally tested.

After the submission of this paper we have noticed some papers 
\cite{berto} which reach similar conclusions.

\section{Acknowledgements}

We are grateful to V. Berezinsky and G. Di Carlo for very useful 
and stimulating discussions. The work of P.B. was funded  
by the DOE and the NASA grant NAG 5-7092 at Fermilab.

\newpage

\begin{table}
\begin{center}
\begin{tabular}{c c c c}
\ ~ & Infrared & Microwave & Radio \\
\hline
\ $I_+$ & $\approx 0.73$ & $0.06$ & $5\cdot 10^{-7}$ \\
\ $I_-$ & No solution  & No solution & No solution \\
\ $II_+$ & $\approx 1$ & $\approx 1$ & $2\times 10^{-3}$ \\
\ $II_-$ & $\approx 1$ & $\approx 1$ & No solution \\
\end{tabular}
\caption{Values of $x$ that solve the equation for the threshold for
pair production in the non-Lorentz invariant approach.}
\label{Table1}
\end{center}
\end{table}

\begin{table}
\begin{center}
\begin{tabular}{c c c c}
\ ~ & Infrared & Microwave & Radio \\
\hline
\ $I_{+}$ & $M\simgt 0.2 M_P$ & $M\simgt 800 M_P$ & $M\simgt 2.5\times 
10^{18} M_P$  \\
\ $I_{-}$ & $M\simgt 6 M_P$ & $M\simgt 3\times 10^4 M_P$ & 
$M\simgt 8\times 10^{19} M_P$  \\
\ $II_+$ & $(M\simgt 3\times 10^{-8} M_P)$ & $(M\simgt 7\times 10^{-6} M_P)$ & 
$ M\simgt 10^{5}M_P $ \\
\ $II_-$ & $(M\simgt 3\times 10^{-7} M_P)$ & $(M\simgt 10^{-4}M_P)$ & 
$M\simgt 10^{6} M_P$ \\
\end{tabular}
\caption{Limits on the scale $M$ where the LI is broken.}
\label{Table2}
\end{center}
\end{table}

\begin{table}
\begin{center}
\begin{tabular}{c c c}
\ ~ & $x$ & Limit\\
\hline
\ $I_+$ & $2 \times 10^{-5}$ & $M\simgt 3\times 10^{13} M_P$\\
\ $I_-$ & No solution & $M\simgt 10^{15} M_P$\\
\ $II_+$ & $0.02$ & $M\simgt 536 M_P$\\
\ $II_-$ & No solution & $M\simgt 6\times 10^3 M_P$\\
\end{tabular}
\caption{Values of $x$ that solve the equations for the threshold of
photopion production on the microwave background (first column) and
lower limit on the scale $M$ of breaking of LI (second column).}
\label{Table3}
\end{center}
\end{table}


\begin{thebibliography}{99}

\bibitem{colgla}
S. Coleman, S.L. Glashow, Phys. Rev. {\bf D59}, 116008 (1999).

\bibitem{wheeler57}
J.A. Wheeler, Annals of Physics {\bf 2}, 604 (1957).

\bibitem{niki}
A.I. Nikishov, Sov. Phys. - JETP {\bf 14}, 393 (1962).

\bibitem{gold}
P. Goldreich and P. Morrison, Sov. Phys. - JETP {\bf 18}, 239 (1964).

\bibitem{gould}
R.J. Gould and G.P. Schreder, Phys. Rev. Lett. {\bf 16}, 252 (1966).

\bibitem{greis}
K. Greisen, Phys. Rev. Lett. {\bf 16}, 748 (1966).

\bibitem{kz}
G.T. Zatsepin and V.A. Kuzmin, Pis'ma Zh. Ekps. Teor. Fiz. {\bf 4}, 114  
(1966) [JETP Lett. {\bf 4}, 78 (1966)].

\bibitem{kir}
D.A. Kirzhnits and V.A. Chechin, Sov. Jour. Nucl. Phys. {\bf 15}, 585 (1971).

\bibitem{mav1}
For a similar approach, not directly related to VHE and UHE Cosmic Rays see: \\
G. Amelino Camelia, J. Ellis, N.E. Mavromatos and S. Sarkar
Nature {\bf 393} (1998) 763.
G. Amelino Camelia, J. Ellis, N.E. Mavromatos and D.V. Nanopoulos
Int. J. Mod. Phys. {\bf A12} (1997) 607.

\bibitem{lgm}
L. Gonzalez-Mestres, Proc. 26th ICRC (Salt Lake City, USA), {\bf 1}, 179 
(1999).

\bibitem{klu}
W. Kluzniak, Astropart. Phys. {\bf 11}, 117 (1999).

\bibitem{kifu}
T. Kifune, Astrophys. J. Lett. {\bf 518} 21 (1999).

\bibitem{bog}
G.Yu. Bogoslovsky and H.F. Goenner, preprint gr-qc/9904081.

\bibitem{ell1}
J. Ellis, N.E. Mavromatos and D.V. Nanopoulos, preprint
hep-th/9909085.

\bibitem{mav2}
J. Ellis, K. Farakos, N.E. Mavromatos, V. Mitsou and D.V. Nanopoulos, 
preprint astro-ph/9907340.

\bibitem{loop}
R. Gambini and J. Pullin, Phys. Rev. {\bf D59} 124021 (1999).

\bibitem{russo}
G. Yu. Bogoslovsky, Fortschr. Physik {\bf 42}, 143 (1994).

\bibitem{garay}
J.L. Garay, Int. J. Mod. Phys. {\bf A14} 4079 (1999). 

\bibitem{ford}
L.H. Ford and Hongwei Yu, preprint gr-qc/9907037. 

\bibitem{luk1}
J. Lukierski, preprint hep-th/9812063.

\bibitem{luk4}
P. Kosinski, J. Lukierski and P. Maslanka, preprint hep-th/9902037.

\bibitem{elltev}
J. Ellis, N.E. Mavromatos and D.V. Nanopoulos, 
Phys. Rev. {\bf D61} 027503 (2000).

\bibitem{tilting}
G. Dvali and M. Shifman, preprint hep-th/9904021, 
{\it To appear in L.B. Okun Festschrift, Eds. V. Telegdi and K. Winter,
to be published by North Olland}.

\bibitem{run}
H. Rund, ``The Differential Geometry on Finsler Spaces'', Berlin 1959.

\bibitem{biller}
S.D. Biller et al., Phys. Rev. Lett. {\bf 83}, 2108 (1999).

\bibitem{whipple}
F. Krennrich et al., Astrophys. J. {\bf 511}, 149 (1999).

\bibitem{hegra}
F.A. Aharonian et al., Astron. \& Astroph. {\bf 349}, 11 (1999).

\bibitem{vvv}
V.V. Vassiliev, preprint astro-ph/9908088.

\bibitem{stecker}
F.W. Stecker, preprint astro-ph/9904416.

\bibitem{coppia}
P.S. Coppi and F.A. Aharonian, Astropart. Phys. {\bf 11}, 35 (1999).

\bibitem{casa}
M. Catanese et al., Astrophys. J. {\bf 469}, 572 (1996).

\bibitem{eastop}
P.L. Ghia et al., Nucl. Phys. B (Proc. Suppl.) {\bf 70}, 506 (1999).

\bibitem{bbv}
V.S. Berezinsky, P. Blasi and A. Vilenkin, Phys. Rev. {\bf D58}, 103515
(1998).

\bibitem{pb}
R.J. Protheroe and P.L. Biermann, Astropart. Phys. {\bf 6}, 45 (1996); 
Erratum-ibid {\bf 7}, 181 (1997).

\bibitem{tak}
M. Takeda et al., Astrophys. J. {\bf 522}, 225 (1999).

\bibitem{hires}
T. Abu-Zayyad et al., Proc. 26th ICRC (Salt Lake City, USA), {\bf 3}, 264
(1999).

\bibitem{hiresidea}
S.C. Corbato et al., Nucl. Phys. B (Proc. Suppl.) {\bf 28B}, 36 (1992).

\bibitem{auger}
J.W. Cronin, Nucl. Phys. B (Proc. Suppl.) {\bf 28B}, 213 (1992).

\bibitem{berto}
O. Bertolami, preprint gr-qc/0001097.

\end{thebibliography}
\end{document}